# CDAS: A Crowdsourcing Data Analytics System

Xuan Liu[†], Meiyu Lu[†], Beng Chin Ooi[†], Yanyan Shen[†], Sai Wu[§], Meihui Zhang[†]
[†]School of Computing, National University of Singapore, Singapore
[§]College of Computer Science, Zhejiang University, Hangzhou, P.R. China

[†]{liuxuan, lumeiyu, ooibc, shenyanyan, mhzhang}@comp.nus.edu.sg, [§]wusai@zju.edu.cn

## ABSTRACT

Some complex problems, such as image tagging and natural language processing, are very challenging for computers, where even state-of-the-art technology is yet able to provide satisfactory accuracy. Therefore, rather than relying solely on developing new and better algorithms to handle such tasks, we look to the crowdsourcing solution – employing human participation – to make good the shortfall in current technology. Crowdsourcing is a good supplement to many computer tasks. A complex job may be divided into computer-oriented tasks and human-oriented tasks, which are then assigned to machines and humans respectively.

To leverage the power of crowdsourcing, we design and implement a Crowdsourcing Data Analytics System, CDAS. CDAS is a framework designed to support the deployment of various crowdsourcing applications. The core part of CDAS is a quality-sensitive answering model, which guides the crowdsourcing engine to process and monitor the human tasks. In this paper, we introduce the principles of our quality-sensitive model. To satisfy user required accuracy, the model guides the crowdsourcing query engine for the design and processing of the corresponding crowdsourcing jobs. It provides an estimated accuracy for each generated result based on the human workers' historical performances. When verifying the quality of the result, the model employs an online strategy to reduce waiting time. To show the effectiveness of the model, we implement and deploy two analytics jobs on CDAS, a twitter sentiment analytics job and an image tagging job. We use real Twitter and Flickr data as our queries respectively. We compare our approaches with state-of-the-art classification and image annotation techniques. The results show that the human-assisted methods can indeed achieve a much higher accuracy. By embedding the quality-sensitive model into crowdsourcing query engine, we effectively reduce the processing cost while maintaining the required query answer quality.

## 1. INTRODUCTION

Crowdsourcing is widely adopted in Web 2.0 sites. For example, Wikipedia benefits from thousands of subscribers, who continually write and edit articles for the site. Another example is

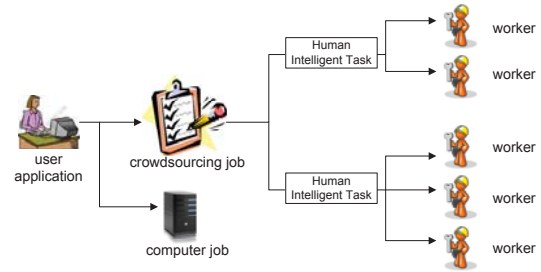

**Figure 1: Crowdsourcing Application**

Yahoo! Answers, where users submit and answer questions. In Web 2.0 sites, most of the contents are created by individual users, not service providers. Crowdsourcing is the driving force of these web sites. To facilitate the development of crowdsourcing applications, Amazon provides the Mechanical Turk (AMT)[1] platform. Computer programmers can exploit AMT's API to publish jobs for human workers, who are good at some complex jobs, such as image tagging and natural language processing. The collective intelligence helps solve many computationally difficult tasks, thereby improving the quality of output and users' experience. Figure 1 illustrates the idea of using crowdsourcing techniques to divide up jobs. CrowdDB [6], HumanGS [19] and CrowdSearch [23] are recent examples of applications on Amazon's AMT crowdsourcing platform.

Crowdsourcing relies on human workers to complete a job, but humans are prone to errors, which can make the results of crowdsourcing arbitrarily bad. The reason is two-fold. First, to obtain rewards, a malicious worker can submit random answers to all questions. This can significantly degrade the quality of the results. Second, for a complex job, the worker may lack the required knowledge for handling it. As a result, an incorrect answer may be provided. To address the above problems, in AMT, a job is split into many HITs (Human Intelligence Tasks) and each HIT is assigned to multiple workers so that replicated answers are obtained. If conflicting answers are observed, the system will compare the answers of different workers and determine the correct one. For example, in the CrowdDB [6], the voting strategy is adopted.

The replication strategy, however, does not fully solve the answer diversity problem. Suppose we want the precision of our image tags to be 95% and the cost of worker per HIT is $0.01. If we assign each HIT to too many workers, we will have to pay a high cost. On the other hand, if few workers provide tags, we will not have enough clue to infer the correct tags. Given an expected accuracy, we therefore need an adaptive query engine that guarantees high accuracy with high probability and incurs as little cost as possible.

---

[1]https://www.mturk.com/mturk/welcome

In this paper, we propose a quality-sensitive answering model for the crowdsourcing systems, which is designed to significantly improve the quality of query results and effectively reduce the processing cost at the same time. This model is the core of our proposed Crowdsourcing Data Analytics System (CDAS). CDAS exploits the crowd intelligence to improve the performance of different data analytics jobs, such as image tagging and sentiment analysis. CDAS transforms the analytics jobs into human jobs and computer jobs, which are then processed by different modules. The human jobs are handled by the crowdsourcing engine, which adopts a two-phase processing strategy. The quality-sensitive answering model is correspondingly split into two sub-models, a prediction model and a verification model. The sub-models are applied to different phases, respectively.

In the first phase, the engine employs the prediction model to estimate how many workers are required to achieve a specific accuracy. The model generates its estimation by collecting the distribution of all workers' historical performances. Based on the model's result, the engine creates and submits the HIT to the crowdsourcing platform. In the second phase, the engine obtains the answers from the human workers and refines them as different workers may return different results for the same question. To verify the answers from different human workers, the voting strategy is used in CrowdDB to select the correct one. In the simplest case, each HIT is sent to $n$ workers ($n$ is odd). A result is assumed to be "correct" and accepted, if no less than $\lceil \frac{n}{2} \rceil$ workers return it. The voting strategy is simple, but is not very effective in the crowdsourcing scenario. Suppose we have a set of product reviews and want to know the opinion of each review. We set the score to either "positive", "negative" or "neutral". If 30% of the workers vote "positive", 30% of the workers vote "negative" and the remaining workers vote "neutral", the voting strategy cannot decide which answer is more trustable. Moreover, even if more than 50% of the workers vote "negative", we cannot accept the answer directly – some malicious workers may collude to produce a false answer. To improve the accuracy of the crowdsourcing results, CDAS adopts a probabilistic approach.

First, a verification model is employed to replace the voting strategy. It relies on workers' past performances (i.e., the workers' accuracies for historical queries) and combines vote distribution and workers' performances. Intuitively, the system is more likely to accept the answers provided by the worker with a good accuracy. A random sampling approach is designed to estimate the workers' accuracies in each job. By applying the probability-based verification model, we can significantly improve the result quality.

Second, instead of waiting for all the results, the adaptive query engine provides an approximate result with confidence and refines it gradually as more answers are returned. This technique has been designed based on our observation that in AMT, workers finish their jobs asynchronously. Therefore, it is important to offer the option of an approximate answer that is gradually improved as more results are available, instead of letting the user wait for the completion of the query. This strategy is similar to the traditional online query processing in philosophy and serves to improve users' experience.

To evaluate our model and the performance of CDAS, we implement two practical crowdsourcing jobs, a twitter sentiment analytics (TSA) job and an image tagging (IT) job. In TSA job, we submit a set of movie titles as our queries and try to find the opinions of Twitter users. In IT job, we use the images of Flickr as the queries and ask the human workers to choose the correct tags. We will show the effectiveness of our crowdsourcing engine based on the quality-sensitive answering model in the experimental section.

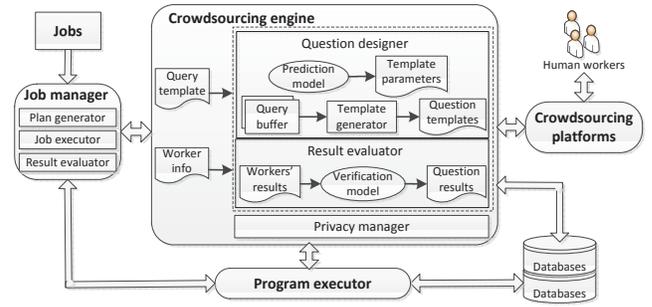

**Figure 2: CDAS Architecture**

The remainder of the paper is organized as follows. In Section 2, we present the architecture of CDAS, and introduce the applications implemented over CDAS. In Section 3, we introduce our prediction model for estimating a proper number of workers for each job. To improve the result accuracy, a probability-based verification model is proposed in Section 4, which can be extended to support online processing. We evaluate the performance of our models in CDAS in Section 5, and discuss some related work in Section 6. We conclude the paper in Section 7.

## 2. OVERVIEW

In this section, we introduce the architecture of our Crowdsourcing Data Analytics System, CDAS, and discuss how to implement applications on top of CDAS.

### 2.1 Architecture of CDAS

CDAS is the system that exploits the crowdsourcing techniques to improve the performance of data analytics jobs. The core difference between CDAS and the conventional analytics systems lies in the processing mechanism. CDAS employs human workers to assist the analytics tasks, while other systems rely solely on computer systems to answer the queries. Figure 2 shows the architecture of CDAS. CDAS consists of three major components: *job manager, crowdsourcing engine* and *program executor*. The job manager accepts the submitted analytics jobs and transforms them into a processing plan, which describes how the other two components (crowdsourcing engine and program executor) should collaborate for the job. In particular, the job manager partitions the job into two parts, one for the computers and one for the human workers. For example, in human-assisted image search, the human workers are responsible for providing the tags for each image, while the image classification and index construction are handled by the computer programs. In most cases, the two parts interact with each other during processing. The program executor summarizes the results of crowdsourcing engine, and the engine may change its job schedule due to the requests of program executor.

The crowdsourcing engine processes human jobs in two phases.

1. In the first phase, the engine generates a query template for the specific type of human jobs. The query template follows the format of the crowdsourcing platform, such as AMT, and should be easily understood by human workers. The engine then translates each job from the job manager into a set of crowdsourcing tasks and publishes them into the crowdsourcing platform. To reduce the crowdsourcing cost, the engine employs a prediction model, which estimates the number of required human workers for a specific task based on the distribution of workers' performance.

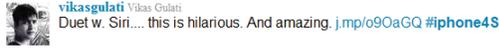

**Figure 3: Query Template**

2. In the second phase, the human workers' answers are returned to the crowdsourcing engine, which combines the results and removes the ambiguity. A verification model is developed to select the correct answer based on the probability estimation.

Sometimes, the human tasks need to disclose some sensitive data to the public. We design a privacy manager inside the engine to address the problem. The privacy manager may adaptively change the formats of the generated questions for human workers. It may also reject some workers for a specific task.

The performance of crowdsourcing engine is determined by the two models, the prediction model and verification model. We shall introduce the two models in the following sections and discuss the implementation of two practical applications, a twitter sentiment analytics (TSA) job and an image tagging (IT) job, to validate the performance of our models.

## 2.2 Deploying Applications on CDAS

In this section, we use the TSA job as a running example to show how to deploy an application on CDAS. TSA job is typically processed using machine learning and information retrieval techniques [2][22]. However, as shown in the experimental section, CDAS can achieve a much higher accuracy than some of these traditional approaches for the TSA job.

In the TSA job, the query is formally defined as follows.

DEFINITION 1. **Query in TSA**
*The query in TSA follows the format of $(S, C, R, t, w)$, where $S$ is a set of keywords, $C$ denotes the required accuracy, $R$ is the domain of answers, $t$ is the timestamp of the query and $w$ is the time window of the query.*

For example, suppose the user wants to know the public opinions for *iPhone4S* from *Oct-14-2011* to *Oct-23-2011*, the corresponding query can be expressed as: $Q$=({*iPhone4S, iPhone 4S*}, 95%, {*Best Ever, Good, Not Satisfied*}, Oct-14-2011, 10). The answer to the query consists of two parts. The first part is the percentage of each opinion and the second part comprises the reasons. For the above query, one possible answer is that most people perceive *iPhone4S* is a good product thanks to the features of Siri and iOS 5, while a smaller but significant number of people are not satisfied with its display and battery performance.

**Table 1: Users' Opinion on iPhone4S**

| Opinions | Percentages | Reasons |
|---|---|---|
| Best Ever | 60% | Siri, iOS 5, Performance |
| Good | 10% | Siri, 1080P |
| Not Satisfied | 30% | iPhone4, Display, Battery |

The query definition of TSA is registered in the job manager, which then generates the corresponding processing plan. The program executor is responsible for retrieving the twitter stream and checking whether the query keyword ($S = iPhone4S$ in above example) exists in a tweet. The candidate tweets are fed to the crowdsourcing engine, which will generate a query template as shown in Figure 3.

When the crowdsourcing engine collects enough tweets in its buffer, it starts to generate the HIT (Human Intelligence Task). In particular, it creates an HTML section (bounded by <div> and </div>) for each tweet using the query's template. For all the tweets in the buffer, we concatenate their HTML sections to form our HIT description. Therefore, one HIT in the TSA job contains questions for multiple tweets about the same product, movie, person or event.

The HIT is then published into the AMT for processing. Algorithm 1 summarizes the two-phase query processing in the crowdsourcing engine (note that Algorithm 1 describes the general query processing strategy, not just for the TSA job). In the preprocessing, the engine generates a HIT job for the tweets using the query template (line 1-6). In the first phase, it applies the prediction model to estimate the number of workers required to satisfy the predefined accuracy (line 7). In the second phase, it submits the HIT to AMT and waits for the answers (line 8-10). The verification model is used to select the correct answers. In line 7, $Q.C$ denotes the accuracy requirement specified by query $Q$.

**Algorithm 1** queryProcessing ( ArrayList<$Tweet$> $buffer$, Query $Q$ )
1: HtmlDesc $H$= new HtmlDesc()
2: **for** $i = 0$ to $buffer$.size-1 **do**
3:     Tweet $t = buffer$.get($i$)
4:     HtmlSection $hs$ = new HtmlSection($Q$.template(), $t$)
5:     $H$.concatenate($hs$)
6: HIT $task$ = new HIT($H$)
7: int $n$=predictWorkerNumber($Q.C$)
8: submit($task, n$)
9: **while** not all answers received **do**
10:     verifyAnswer()

In Algorithm 1, the two models direct the whole procedure of query processing, which are also the focus of this paper and will be presented in the following sections.

## 3. PREDICTION MODEL

### 3.1 Economic Model in AMT

The prediction model is designed to ensure high-quality answers and to reduce cost. It is highly related to how the crowdsourcing platform charges the requesters. Therefore, we first briefly introduce the economic model of AMT.

In AMT, a HIT is published and broadcasted to all candidate workers. Any candidate worker can accept the task. Thus, if $n$ answers for a HIT are required, from the point of view of CDAS, there will be $n$ random workers providing the answers. AMT charges CDAS for each HIT using the following rules:

1. Every worker is paid a fixed amount of money $m_c$.

2. CDAS pays a fixed amount of money $m_s$ per worker to the AMT system for each HIT.

Therefore, we spend $(m_c + m_s)n$ for each HIT. Take query $Q = (S, C, R, t, w)$ in TSA as an example, if we get $K$ available tweets

1042

for each time unit, the cost of processing $Q$ is $(m_c+m_s)nKw$. In our predication model, the number of workers is correlated to the required accuracy $C$. We use function $g$ to denote the relationship between $C$ and $n$. Consequently, the query cost can be represented as $(m_c+m_s)wK \times g(C)$. Before we present the technical details, we summarize the notations used in the paper in Table 2.

**Table 2: Table of Notations**

| | |
|---|---|
| $\mathcal{U}$ | the set of workers |
| $u_i$ | the $i$-th worker |
| $n$ | the number of workers |
| $P_{\frac{n}{2}}$ | the probability of at least $\lceil \frac{n}{2} \rceil$ workers provide the correct answer |
| $\mathcal{A}$ | the set of accuracy of workers |
| $a_i$ | the accuracy of worker $u_i$ |
| $\mu$ | the mean value of worker accuracy |
| $f(u_i)$ | the answer provided by worker $u_i$ |
| $\Omega$ | the observation of distribution of answers |
| $P(\overline{r}|\Omega)$ | the probability of answer $\overline{r}$ being correct under the observation $\Omega$ |
| $m$ | the number of all possible answers |
| $c_i$ | the confidence of worker $u_i$ |
| $\rho(r_i)$ | the confidence of answer $r_i$ |

## 3.2 Voting-based Prediction

Given $n$ ($n$ is odd) answers from workers $\mathcal{U} = \{u_1, u_2, ..., u_n\}$, the voting strategy accepts an answer if at least $\lceil \frac{n}{2} \rceil$ workers return the same answer. While the voting strategy guarantees that no other answers have more votes of being the correct answer, it however does not address the problem of how to select $n$.

To address the above problem, we propose a voting-based prediction model. Given an accuracy requirement, the prediction model estimates the number of workers required. That is, the goal of the prediction model is to derive the function $g$ for each query. We prove in Section 4 that the model can also produce a bound for our probability-based verification approach.

### 3.2.1 A Conservative Estimation

We compute the probability that at least $\lceil \frac{n}{2} \rceil$ workers provide the correct answer. We use $P_{\frac{n}{2}}$ to denote the probability. Suppose the accuracy of all $n$ workers are $\mathcal{A} = \{a_1, a_2, \cdots, a_n\}$, where the accuracy means the probability of a worker providing a correct answer. By the definition of $P_{\frac{n}{2}}$, we have the following equation:

$$P_{\frac{n}{2}} = \sum_{\mathbb{U} \subseteq \mathcal{U}, |\mathbb{U}| \geq \lceil \frac{n}{2} \rceil} (\prod_{u_i \in \mathbb{U}} a_i \prod_{u_j \notin \mathbb{U}} (1-a_j))$$

$\mathbb{U}$ denotes a subset of user set $\mathcal{U}$ with size no smaller than $\lceil \frac{n}{2} \rceil$. The above equation enumerates all the possible cases that the correct answer can be obtained by voting.

The workers of a HIT can be considered as random workers from AMT. Let $\mu$ denote the mean value of the workers' accuracy. We have the following theorem to compute the expectation of the probability that at least $\lceil \frac{n}{2} \rceil$ workers return the correct answer:

THEOREM 1. *If workers answer the queries independently,*

$$E[P_{\frac{n}{2}}] = \sum_{k=\lceil \frac{n}{2} \rceil}^{n} \binom{n}{k} \mu^k (1-\mu)^{n-k}$$

PROOF. As all workers are randomly picked, $a_i$ and $a_j$ are independent for any $i \neq j$. Similarly, $a_i$ and $1-a_j$ are also independent. Thus,

$$\begin{aligned} E[P_{\frac{n}{2}}] &= E[\sum_{\mathbb{U} \subset \mathcal{U}, |\mathbb{U}| \geq \lceil \frac{n}{2} \rceil} (\prod_{u_i \in \mathbb{U}} a_i \prod_{u_j \notin \mathbb{U}} (1-a_j))] \\ &= E[\sum_{k=\lceil \frac{n}{2} \rceil}^{n} (\sum_{\mathbb{U} \subset \mathcal{U}, |\mathbb{U}|=k} (\prod_{u_i \in \mathbb{U}} a_i \prod_{u_j \notin \mathbb{U}} (1-a_j)))] \\ &= \sum_{k=\lceil \frac{n}{2} \rceil}^{n} (\sum_{\mathbb{U} \subset \mathcal{U}, |\mathbb{U}|=k} (\prod_{u_i \in \mathbb{U}} E[a_i] \prod_{u_j \notin \mathbb{U}} E[(1-a_j)])) \end{aligned}$$

We have $E[a_i] = \mu$ and $E[1-a_i] = 1-\mu$. Therefore, $E[P_{\frac{n}{2}}]$ can be computed as:

$$\begin{aligned} E[P_{\frac{n}{2}}] &= \sum_{k=\lceil \frac{n}{2} \rceil}^{n} (\sum_{\mathbb{U} \subset \mathcal{U}, |\mathbb{U}|=k} (\prod_{u_i \in \mathbb{U}} \mu \prod_{u_j \notin \mathbb{U}} (1-\mu))) \\ &= \sum_{k=\lceil \frac{n}{2} \rceil}^{n} (\sum_{\mathbb{U} \subset \mathcal{U}, |\mathbb{U}|=k} \mu^k (1-\mu)^{n-k}) \\ &= \sum_{k=\lceil \frac{n}{2} \rceil}^{n} \binom{n}{k} \mu^k (1-\mu)^{n-k} \end{aligned}$$

□

For a given query, we require $E(P_{\frac{n}{2}})$ to be no less than a given accuracy $C$, i.e., $E(P_{\frac{n}{2}}) \geq C$. Furthermore, we derive a lower bound of $E(P_{\frac{n}{2}})$ that can be easily computed as follows.

THEOREM 2. $E[P_{\frac{n}{2}}] \geq 1 - e^{-2n(\mu-\frac{1}{2})^2}$

PROOF. By Chernoff Bound,

$$\sum_{k=\lfloor \frac{n}{2} \rfloor +1}^{n} \binom{n}{k} \mu^k (1-\mu)^{n-k} \geq 1 - e^{-2n(\mu-\frac{1}{2})^2}$$

Moreover, for any odd $n$, we have

$$\lfloor \frac{n}{2} \rfloor + 1 = \lceil \frac{n}{2} \rceil$$

Therefore,

$$\begin{aligned} E[P_{\frac{n}{2}}] &= \sum_{k=\lceil \frac{n}{2} \rceil}^{n} \binom{n}{k} \mu^k (1-\mu)^{n-k} \\ &= \sum_{k=\lfloor \frac{n}{2} \rfloor +1}^{n} \binom{n}{k} \mu^k (1-\mu)^{n-k} \geq 1 - e^{-2n(\mu-\frac{1}{2})^2} \end{aligned}$$

□

By requiring $1 - e^{-2n(\mu-\frac{1}{2})^2} \geq C$, we guarantee that $E[P_{\frac{n}{2}}] \geq C$ (i.e., the expected accuracy of the query result is no less than $C$). Consequently, we obtain a sufficient condition for the quality of the crowdsourcing query engine:

THEOREM 3. *Given required accuracy $C$ and the mean value of workers' accuracy $\mu$, choosing*

$$n \geq \frac{-\ln(1-C)}{2(\mu-\frac{1}{2})^2}$$

*workers ensures the expected accuracy of the crowdsourcing result no less than $C$.*

Note that $n$ is an odd integer, so the minimum value of $n$ is $2\lfloor \frac{-\ln(1-C)}{4(\mu-\frac{1}{2})^2} \rfloor + 1$.

### 3.2.2 Optimization with Binary Search

Setting $n$ to $2\lfloor \frac{-\ln(1-C)}{4(\mu-\frac{1}{2})^2} \rfloor + 1$ ensures the expected accuracy of results. However, it is well known that Chernoff Bound provides a tight estimation only for a large enough $n$. In some HITs, only a few workers participate in processing. Therefore, Theorem 3 generates a conservative estimation that may cause too many workers to be involved. To address this problem, we use Theorem 3 as an upper bound and apply a binary search algorithm (on odd numbers) to find a tighter estimation, i.e. the minimum odd $n$ that satisfies $E[P_{\frac{n}{2}}] \geqslant C$.

---

**Algorithm 2** binarySearch(double $C$)

//$C$ is the required accuracy
1: int $s = 1$, int $e = 2\lfloor \frac{-\ln(1-C)}{4(\mu-\frac{1}{2})^2} \rfloor + 1$
2: **while** $s < e$ **do**
3:     int $m = 2\lfloor \frac{s+e}{4} + \frac{1}{2} \rfloor - 1$
4:     int $E_m$=computeExpectedProb($m$)
5:     **if** $E_m \geqslant C$ **then**
6:        $e = m$
7:     **else**
8:        $s = m + 2$
9: **return** $e$

---

**Algorithm 3** computeExpectedProb(int $x$)

1: double $E$=0, $\delta = \mu^x$
2: **for** int $i$=x to $\lceil \frac{x}{2} \rceil$ **do**
3:     $E = E + \delta$
4:     $\delta = \delta \times \frac{(1-\mu)i}{\mu(x-i+1)}$
5: **return** $E$

---

Algorithm 2 shows the idea of binary search. We initialize the domain of $n$ to be $[1, 2\lfloor \frac{-\ln(1-C)}{4(\mu-\frac{1}{2})^2} \rfloor + 1]$ (line 1). At each step, we compute the expected accuracy of using $m$ workers (line 4), until we reach the minimum $m$ that satisfies the accuracy requirement. Algorithm 3 illustrates the process of computing the expected accuracy. Its correctness is based on the fact that $\binom{n}{k-1}/\binom{n}{k} = k/(n-k+1)$. Obviously, the time complexity of Algorithm 3 is $O(n)$. Therefore, we can get a tighter bound of the number of workers required using Algorithm 2 in $O(n \log n)$ time.

### 3.3 Sampling-based Accuracy Estimation

In the previous two prediction models, we rely on the statistics of workers' accuracy distribution. However, not all crowdsourcing platforms provide such information due to the privacy issue. Even if some platforms provide certain statistics, they cannot be directly used as workers' accuracy. For example, AMT system records the approval rate of each worker. Approval rate shows the percentage of answers approved by the requester. However, we have observed that the approval rate is not consistent with the accuracy of the worker in CDAS. There are two main reasons. First, the worker's accuracy may vary widely across jobs. Second, some requesters set automatic approval for all answers without verification. The difference of approval rate and accuracy is studied through experiments. To resolve the above problem, we design a sampling-based approach. Specifically, for a registered query, we randomly embed $m$ questions, whose ground truth are known beforehand. These questions are used as our testing samples to estimate the workers' accuracy.

Here we use TSA application to illustrate the sampling method. As mentioned previously, each HIT contains the questions of $\mathcal{B}$ tweets. To get unbiased results, we randomly inject $\alpha\mathcal{B}$ samples

---

**Algorithm 4** doSampling(HIT $H$)

1: WorkerSet $U$=$H$.getWorkers()
2: Double[] $rate$ = new Double[$U$.size]
3: **while** $H$.nextQuestion()$\neq$ null **do**
4:     Question $q = H$.getNextQuestion()
5:     **if** $q$ is a testing sample **then**
6:        **for** $i = 0$ to $U$.size **do**
7:           Worker $u = U$.get($i$)
8:           **if** $u$.getAnswer($q$)==$q$.groundTruth **then**
9:              $rate[i] = rate[i] + \frac{1}{\alpha\mathcal{B}.size}$

---

into a HIT. In other words, each HIT has $\alpha\mathcal{B}$ testing samples and $(1-\alpha)\mathcal{B}$ new tweets. In our current implementation, $\alpha$ and $\mathcal{B}$ are set to 0.2 and 100, respectively. We evaluate the effect of sampling rate $\alpha$ in our experiments, and the results confirm that even a low sampling rate can produce an acceptable estimation.

In the sampling process, CDAS collects the accuracy of participating workers. Algorithm 4 shows the procedure. After the sampling, the statistics are used in both the prediction model and the verification model.

## 4. VERIFICATION MODEL

In the voting-based verification, if more than half of the workers return the same answer, the query engine will accept it as the correct answer. Despite the fact that our predication model tries to guarantee that at least half of the workers submit the correct answer, the voting-based verification occasionally fails to provide an answer.

For a specific question, different workers may provide different answers, and in some cases, no answer gets an agreement above 50%. Moreover, the voting strategy assumes that all the workers provide the correct answer with the same probability, which is not true as the accuracy of different workers varies a lot and the workers with higher accuracy are more trustable. In this section, we propose a probability-based verification method to determine the best answer.

### 4.1 Probability-based Verification

Probability-based verification tries to evaluate the quality of answers through workers' historical performances (i.e. accuracy). In particular, given the probability distribution of workers' performances, we apply the Bayesian theorem to estimate the accuracy of each result. We adopt and extend the approach proposed in the data fusion [4] for integrating conflicting results in the CDAS.

Suppose a HIT is answered by $n$ workers $\{u_1, u_2, \cdots, u_n\}$ with accuracy $\{a_1, a_2, \cdots, a_n\}$. We define function $f(u_i)$ to represent the answer provided by worker $u_i$. Based on Bayesian analysis, the probability of a specific answer $\bar{r} \in R$ being the correct answer given the observation of the answer's distribution $\Omega$ (i.e. the answers provided by $n$ workers) can be computed as:

$$
\begin{aligned}
P(\bar{r}|\Omega) &= \frac{P(\Omega|\bar{r})P(\bar{r})}{P(\Omega)} \\
&= \frac{P(\Omega|\bar{r})P(\bar{r})}{\sum_{r_i \in R} P(\Omega|r_i)P(r_i)}
\end{aligned}
$$

Suppose the size of the answer domain $|R| = m$. Without a priori knowledge, each answer $r_i \in R$ appears with equal probability of $\frac{1}{m}$. Then the above equation can be transformed into:

$$P(\bar{r}|\Omega) = \frac{P(\Omega|\bar{r})}{\sum_{r_i \in R} P(\Omega|r_i)} \quad (1)$$

Let $\bar{r}$ be the correct answer. The probability for worker $u_j$ providing the correct answer is $a_j$ (i.e. accuracy). Without any priori knowledge, each incorrect answer provided by $u_j$ appears with equal probability $\frac{1-a_j}{m-1}$. Therefore, $P(\Omega|\bar{r})$ can be computed as:

$$P(\Omega|\bar{r}) = \prod_{f(u_j)=\bar{r}} a_j \prod_{f(u_j)\neq\bar{r}} \frac{1-a_j}{m-1} \quad (2)$$

Combining Equation 1 and 2, we have

$$P(\bar{r}|\Omega) = \frac{\prod_{f(u_j)=\bar{r}} a_j \prod_{f(u_j)\neq\bar{r}} \frac{1-a_j}{m-1}}{\sum_{r_i \in R}(\prod_{f(u_j)=r_i} a_j \prod_{f(u_j)\neq r_i} \frac{1-a_j}{m-1})}$$

$$= \frac{\prod_{f(u_j)=\bar{r}} \frac{(m-1)a_j}{1-a_j}}{\sum_{r_i \in R}(\prod_{f(u_j)=r_i} \frac{(m-1)a_j}{1-a_j})} \quad (3)$$

For ease of illustration, we define the *Worker Confidence* for an answer as follows.

DEFINITION 2. **Worker Confidence**
Let $a_j$ be the accuracy of worker $u_j$. The confidence $c_j$ of worker $u_j$ is defined as:

$$c_j = \ln\frac{(m-1)a_j}{1-a_j} = \ln(m-1) + \ln\frac{a_j}{1-a_j}$$

From the above definition, we can see that high-accuracy workers will get large confidence values. This is consistent with the intuition that workers with higher accuracy are more trustable.

Based on the definition of worker confidence and the equation 3, we define the *Answer Confidence* as below.

DEFINITION 3. **Answer Confidence**
The confidence of an answer $\bar{r}$ equals to the probability of $\bar{r}$ being the correct answer:

$$\rho(\bar{r}) = P(\bar{r}|\Omega) = \frac{e^{\sum_{f(u_j)=\bar{r}} c_j}}{\sum_{r_i \in R}(e^{\sum_{f(u_j)=r_i} c_j})} \quad (4)$$

In our CDAS, the answer with the highest confidence is accepted as the final result. In fact, the confidence of an answer represents a variant of voting, where $e^{c_j}$ is used as the weight for worker $u_j$. Apparently, the worker with a higher confidence gets more weight. To speed up the computation of $P(\bar{r}|\Omega)$, we cache the value $\ln\frac{a_j}{1-a_j}$ for each known worker.

We can prove that using Theorem 1 to estimate the number of workers required also produces a quality bound for our probability-based verification approach.

THEOREM 4. *If $E[P_{\frac{n}{2}}] \geq C$ and let $\bar{r}$ be the correct answer, we have that our probability-based verification model returns $\bar{r}$ as the result with a probability no less than $C$.*

PROOF. Based on Theorem 1,

$$E[P_{\frac{n}{2}}] = \sum_{k=\lceil \frac{n}{2} \rceil}^{n} \binom{n}{k} \mu^k (1-\mu)^{n-k} \geq C$$

Namely, the expected number of workers, who provide the correct answer, is larger than $\frac{n}{2}$ with a probability larger than $C$. The confidences of all workers are independent and identically distributed (i.i.d.), because the accuracies of the workers are i.i.d. Let $E_c$ denote the mean value of workers' confidences. As a result, the total number of expected votes for answer $\bar{r}$ is

$$E[\sum_{f(u_j)=\bar{r}} c_j] = \sum_{f(u_j)=\bar{r}} E[c_j]$$
$$= E_c \cdot |\{u_j|f(u_j)=\bar{r}\}|$$
$$> \frac{n}{2}E_c$$

Note that in Equation 4, all answers share the same denominator. The value of $P(\bar{r}|\Omega)$ is proportional to $e^{\sum_{f(u_j)=\bar{r}} c_j}$. Thus, $\bar{r}$ is the answer with the largest expected confidence and is returned as the result in expectation. Otherwise, if another answer $r'$ has a larger expected probability than $\bar{r}$, i.e.,

$$E[P(r'|\Omega)] > E[P(\bar{r}|\Omega)]$$

Therefore,

$$E[\sum_{f(u_j)=r'} c_j] > E[\sum_{f(u_j)=\bar{r}} c_j] > \frac{n}{2}E_c$$

We will have

$$E[\sum_{f(u_j)=r'} c_j] + E[\sum_{f(u_j)=\bar{r}} c_j] > nE_c$$

In fact, the sum of workers' confidences is equal to the sum of confidences for every answer:

$$nE_c = E[\sum_{r_i \in R_p} (\sum_{f(u_j)=r_i} c_j)]$$

This results in a contradiction that the sum of confidences of $\bar{r}$ and $r'$ exceeds the sum of all confidences. Therefore, our probability-based verification model returns $\bar{r}$ as the result with a probability no less than $C$. □

The only unknown parameter in Equation 4 is $m$, the size of $R$. We can simply set $m = |R|$. However, in our experimental study, we have found that not all answers in $R$ are picked by the workers. For example, if a question asks a worker to rank a product based on some tweets and the score ranges from 0 to 100, the scores will follow a very skewed distribution. Some low-probability answers are never selected, but they do reduce the weight of a correct answer. Thus, we need to select a good $m$ to prune the noise.

After a HIT completes, the crowdsourcing engine gets $k$ distinct answers for a specific question from $n$ workers ($k \leq n$). In this observation, we select $k$ distinct answers among $m$ possible ones. The probability of this selection can be computed as $\frac{\binom{m}{k}}{m^k}$. Suppose this is not a very rare observation and the probability of this observation is larger than $\epsilon$ (e.g., we prune the low-probability noise). The following lemma provides a lower bound for $m$.

LEMMA 1. $m > \dfrac{k-1}{H_{k-1} - (k-1)(k\epsilon)^{\frac{1}{k-1}}}$, where $H_k = \sum_{i=1}^{k} \frac{1}{i}$ is the k-th Harmonic number.

PROOF.

$$\epsilon < \frac{\binom{m}{k}}{m^k} = \frac{1}{m^k} \frac{m(m-1)\cdots(m-k+1)}{k(k-1)\cdots 1}$$

$$= \frac{1(1-\frac{1}{m})(1-\frac{2}{m})\cdots(1-\frac{k-1}{m})}{k \cdot 1 \cdots (k-1)}$$

$$= \frac{1}{k}(1-\frac{1}{m})(\frac{1}{2}-\frac{1}{m})\cdots(\frac{1}{k-1}-\frac{1}{m})$$

$$\leq \frac{1}{k}(\frac{1}{k-1}(\sum_{i=1}^{k-1}(\frac{1}{i}-\frac{1}{m})))^{k-1}$$

$$= \frac{1}{k}(\frac{H_{k-1}}{k-1}-\frac{1}{m})^{k-1}$$

Derived from the above equation, we have

$$(k\epsilon)^{\frac{1}{k-1}} < \frac{H_{k-1}}{k-1} - \frac{1}{m}$$

Therefore

$$m > \frac{1}{\frac{H_{k-1}}{k-1} - (k\epsilon)^{\frac{1}{k-1}}} = \frac{k-1}{H_{k-1} - (k-1)(k\epsilon)^{\frac{1}{k-1}}}$$

□

For a large $k$, the above lower bound is too loose. Instead, we propose a tighter lower bound for $m$:

LEMMA 2. $m > \dfrac{k-1}{1 - \frac{\epsilon k}{e^k}}$

PROOF. From Lemma 1, we have

$$\epsilon < \frac{\prod_{i=1}^k (1 - \frac{i-1}{m})}{\prod_{i=1}^k i}$$

Therefore,

$$\ln \epsilon < \sum_{i=1}^k \ln \frac{1 - \frac{i-1}{m}}{i}$$

Obviously,

$$\sum_{i=1}^k \ln \frac{1 - \frac{i-1}{m}}{i} > k \ln \frac{1 - \frac{k-1}{m}}{k}$$

By setting $k \ln \dfrac{1 - \frac{k-1}{m}}{k} > \ln \epsilon$, we get a tighter bound:

$$m > \frac{k-1}{1 - \frac{\epsilon k}{e^k}}$$

□

THEOREM 5.

$$m > \max\{\frac{k-1}{H_{k-1} - (k-1)(k\epsilon)^{\frac{1}{k-1}}}, \frac{k-1}{1 - \frac{\epsilon k}{e^k}}\}$$

PROOF. Directly from Lemma 1 and 2. □

In our verification, we set $\epsilon$ to 0.05 based on Fisher's exact test [5], which is widely adopted in practice. We then use Theorem 5 to estimate the value of $m$.

We now give an example in TSA to show the benefit of applying our probability-based verification model. Table 3 shows the example. Five workers with different accuracies provide three different answers, namely *Positive*, *Neutral* and *Negative*. The results of the three verification models are shown in Table 4. Both the *Half-Voting* model and the *Majority-Voting* model choose *Positive* as the results since three workers out of five provide the answer *Positive*. However, our verification model can correctly choose *Negative* as the result because the worker answering *Negative* has a much higher accuracy. As a result, our verification model gets more accurate answers than the other two voting-based models.

Table 3: An Example of Workers' Answers

| Movie Title | Green Latern | | | | |
|---|---|---|---|---|---|
| Tweet | *Oh. My. GOD. "Green Lantern" movie is terrible. Like, "Lost In Space" movie terrible.* | | | | |
| Worker ID | $w_1$ | $w_2$ | $w_3$ | $w_4$ | $w_5$ |
| Accuracy | 0.54 | 0.31 | 0.49 | 0.73 | 0.46 |
| Answer | pos | pos | neu | neg | pos |

Table 4: Results of Verification Models

| | pos | neu | neg | Answer |
|---|---|---|---|---|
| Half-Voting | 3 | 1 | 1 | pos |
| Majority-Voting | 3 | 1 | 1 | pos |
| Verification | 0.329 | 0.176 | 0.495 | neg |

### 4.2 Online Processing

The workers submit their answers asynchronously in the AMT and CDAS has to wait for sufficient number of answers to be submitted. As a consequence, query response time in CDAS (and other crowdsourcing systems for that matter) is expected to be longer than that of non-crowdsourcing systems. To alleviate such a problem and also to improve users' experience, we adopt online processing techniques in CDAS. Instead of waiting for all workers to complete their tasks, CDAS provides an approximate result based on the answers received so far. As we have previously discussed, uncertainty and approximation cannot be avoided in crowdsourcing systems, which makes online processing a perfect fit for the query processing in CDAS.

To resolve the uncertainty, we extend the techniques of data fusion [4][14] to estimate the answer's confidence. However, the same approach cannot be directly applied to the online processing in CDAS, as in the crowdsourcing systems, the human workers compete for the tasks and CDAS does not have the profile (i.e. accuracy) for a specific user until he/she returns the answer. In our case, the accuracy of the answer provided by an unseen worker can only be estimated by the distribution of all workers' accuracies.

#### 4.2.1 Finding the Correct Answer Online

We apply Equation 4 to continuously update the probability of each received answer. Suppose a HIT is assigned to $n$ workers and the query engine receives answers from $n'$ ($n' < n$) workers. Unlike Equation 4, in this case, we only receive a partial observation $\Omega'$ for the answer distribution. For the remaining $n - n'$ workers, we have no idea about what answers they may provide. Let $s$ denote a possible answer set by the remaining workers and we use $\mathcal{S}$ to represent all the possible $s$.

Let $A = \{a_{n'+1}, a_{n'+2}, ..., a_n\}$ be the accuracies of the remaining $n - n'$ workers. As we do not know the identities of the remaining $n - n'$ workers, we consider all the possibilities. We use $\mathbb{A}$ to represent all the possible permutations of $A$. The confidence of an answer $r$ being the correct one can be estimated as the expected probability $P(r|\Omega', s)$ over $\mathcal{S}$ and $\mathbb{A}$, i.e.,

$$\rho(r) = E_{s \in \mathcal{S}, A \in \mathbb{A}}[P(r|\Omega', s)]$$

The following theorem shows that Equation 4 can be applied to compute $\rho(r)$.

THEOREM 6. *Assume that workers process the query independently and the answers are submitted in a random order.* $\rho(r) = P(r|\Omega')$

PROOF. Based on the assumption, we have:

$$\begin{aligned} \rho(r) &= E_{s\in\mathcal{S}, A\in\mathbb{A}}[P(r|\Omega', s)] \\ &= E_{A\in\mathbb{A}}[E_{s\in\mathcal{S}}[P(r|\Omega', s)]] \end{aligned}$$

In fact, the answer set of the remaining workers $s$ does not affect the computation of the above equation. As shown in [14],

$$E_{s\in\mathcal{S}}[P(r|\Omega', s)] = P(r|\Omega')$$

The computation of $\rho(r)$ can be further simplified as:

$$\begin{aligned} \rho(r) &= E_{A\in\mathbb{A}}[E_{s\in\mathcal{S}}[P(r|\Omega', s)]] \\ &= E_{A\in\mathbb{A}}[P(r|\Omega')] \\ &= P(r|\Omega') \end{aligned}$$

□

Theorem 6 shows that the confidence of a partial result can also be computed by Equation 4. Therefore, we select the answer with maximal confidence as our correct answer.

### 4.2.2 Early Termination

When the current answers are good enough, we can terminate the HIT to reduce cost. The major challenge of early termination is how to measure the quality of the current results. Intuitively, we can stop accepting answers from new workers as soon as we are sure that the current result $r$ will not change by the answers we choose to forgo.

In particular, let $r_1$ and $r_2$ be the best and second best answers based on their confidence, respectively. We have $P(r_1|\Omega') > P(r_2|\Omega')$. Let $(u_1, u_2, ..., u_n)$ be the set of workers. Suppose $n'$ workers have submitted answers and $n - n'$ answers remain unfilled. Assume an answer set $s = \{f(u_i) = r_2|n'+1 \leqslant i \leqslant n\}$. Using similar techniques in paper [14], we can prove the theorem of minimal possible value of $P(r_1|\Omega)$ and the maximal possible value of $P(r_2|\Omega)$:

$$\min P(r_1|\Omega) = P(r_1|\Omega', s) \quad (5)$$

$$\max P(r_2|\Omega) = P(r_2|\Omega', s) \quad (6)$$

Note that $\min P(r_1|\Omega)$ and $\max P(r_2|\Omega)$ are related to the random variables $a_{n'+1}, a_{n'+2}, \cdots, a_n$. In our algorithm, we use the expected value of $\min P(r_1|\Omega)$ and $\max P(r_2|\Omega)$, namely, $E_{A\in\mathbb{A}}[\min P(r_1|\Omega)]$ and $E_{A\in\mathbb{A}}[\max P(r_2|\Omega)]$. However, it is difficult to compute the expected values directly. Therefore, in practice, we use the approximate values of $E_{A\in\mathbb{A}}[\min P(r_1|\Omega)]$ and $E_{A\in\mathbb{A}}[\max P(r_2|\Omega)]$. We assume every remaining worker has the same accuracy $E[a_i]$ and use it in the Equation 5 and 6. Empirical results show that the approximations work well in practice.

We propose three different strategies as the termination condition:

**MinMax** $E_{A\in\mathbb{A}}[\min P(r_1|\Omega)] > E_{A\in\mathbb{A}}[\max P(r_2|\Omega)]$

**MinExp** $E_{A\in\mathbb{A}}[\min P(r_1|\Omega)] > P(r_2|\Omega')$

**ExpMax** $P(r_1|\Omega') > E_{A\in\mathbb{A}}[\max P(r_2|\Omega)]$

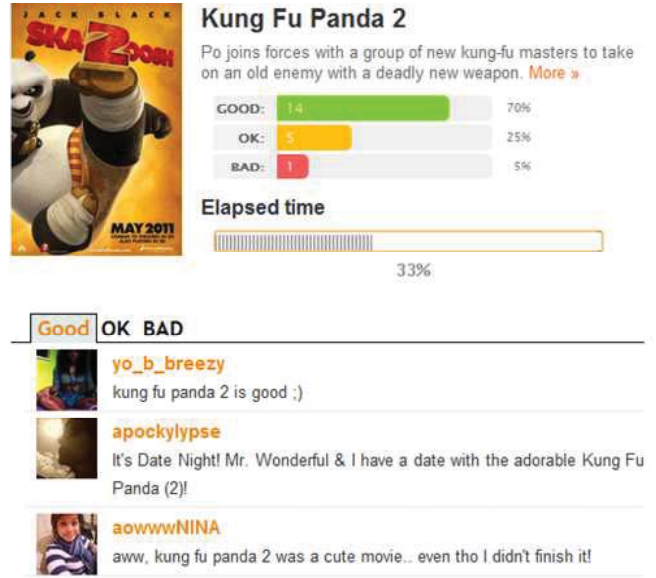

**Figure 4: Reviews for Kung Fu Panda 2**

*MinMax* guarantees that the answer output by our system is stable when the termination condition is achieved. However, it is too conservative. *MinExp* and *ExpMax* can terminate the processing much earlier, but may lead to low-quality results. We study the effect of the three strategies in our experiments.

---

**Algorithm 5** onlineProcessing(Question $q$)
---
1: Set $answer$=new Set()
2: Map$<Answer, float> result$= new Map()
3: **while** not all answers are returned **do**
4:     Answer $A$ = getNextAnswer($q$)
5:     $answer$.add($A$)
6:     Set $distinctAnswer$ = getDistinctAnswer($answer$)
7:     **for** $i = 0$ to $distinctAnswer$.size-1 **do**
8:         Answer $A= distinctAnswer$.get($i$)
9:         float $confidence$ = computeConfidence($A$)
10:        $result$.put($A, confidence$)
11:    **if** canTerminate($result$) **then**
12:        break
13: return $result$

---

Algorithm 5 outlines the online processing strategy adopted in CDAS. The query engine continuously updates the confidence of each answer (line 3-13) until the termination condition is satisfied. We apply Equation 4 to estimate the confidence of each answer (line 9) and apply one of the three termination strategies to decide whether to stop the processing (line 11).

## 4.3 Result Presentation

In the onlineProcessing Algorithm (Algorithm 5), if there is an answer that meets the termination condition, online processing will stop and CDAS will accept the answer. Otherwise, if none of the answers is good enough, CDAS will update the confidence of each answer according to Equation 4.

We take queries in TSA as an example to illustrate the result presentation. Given a list of tweets $t_1, t_2, ..., t_N$, let function $h_{t_i}(r)$ return the score of answer $r$ for tweet $t_i$. $h_{t_i}(r)$ is defined as follows:

$$h_{t_i}(r) = \begin{cases} 1 & \text{if r is accepted for } t_i \\ 0 & \text{if another answer is accepted} \\ \rho_{t_i}(r) & \text{none of the answers are accepted} \end{cases}$$

The percentage of answer $r$ is then computed as $\frac{1}{N}\sum_{i=1}^{N} h_{t_i}(r)$. Moreover, we generate a set of keywords as reasons for each answer $r$. These keywords are the most frequent keywords submitted by the workers who have provided the answer $r$. The results are updated as new tweets are being streamed into TSA.

Figure 4 shows the online processing interface of TSA for the review results of *Kung Fu Panda 2*. It summarizes Twitter users' opinions into three categories. The time window of the query is set to 12 minutes and in the elapsed time (4 minutes), 20 tweets are fed to TSA, among which 70% of tweets say *Kung Fu Panda 2* is a good movie. TSA updates the result upon new tweets arriving. Users can click an answer to expand the view. TSA will list the corresponding tweets for the answer. The tweets are sorted based on timestamps from the newest to the oldest. The user can also check the progress of the current running HIT.

## 5. PERFORMANCE EVALUATION

To evaluate the effectiveness of the quality-sensitive answering model in CDAS, we developed two crowdsourcing applications, a twitter sentiment analytics (TSA) job and an image tagging (IT) job. We present the comprehensive experimental results over TSA, and due to the space constraint, we shall only provide the comparison with an online image tagging toolkit for the IT application. The results for the other experiments over IT exhibit similar trends to those of TSA.

By default, our approach applies the probability-based verification model (denoted as *Verification*) to select the best answer. For comparison, the *Half-Voting* and *Majority-Voting* models are used as two alternative verification approaches. Suppose $n$ ($n$ is odd) workers are employed for a particular task. In the *Half-Voting* model, the answer $r_i$ is accepted only if no less than $\frac{n}{2}$ workers return it as their answers. In the *Majority-Voting* model, let $v(r_i)$ denote the votes for answer $r_i$. The answer $r_i$ is accepted if for any other answer $r_j$, $v(r_i) > v(r_j)$.

## 5.1 Application 1: TSA

We deploy TSA on AMT and use 200 movie titles as our queries. The selected titles are the most recent movies listed in IMDB (Internet Movie Database). The query follows the format of $Q$=({*movie name*}, *accuracy requirement*, {*Positive, Neural, Negative*}, *Oct-1-2011, 1 day*). Namely, the queries are processed against one-day tweets. For each HIT, 30 workers are employed to perform the review categorization task. We manually check each of the reviews to generate our ground truth.

### 5.1.1 Crowdsourcing vs. SVM Algorithm

We first show the advantages of crowdsourcing techniques over computer programs. We compare the results of TSA with LIBSVM[2]. To build an automatic classification model using LIBSVM, tweet reviews about five movies are selected as the test data, and tweets about the rest 195 movies are used as training data. After a stream of tweets passes the filters of TSA, we also send it to LIBSVM and collect the corresponding results. We then compare the results against our ground truth. In TSA, we vary the number of workers from 1 to 5. Figure 5 shows the accuracies of both systems for five movies, each with 200 tweet reviews. In most cases, TSA can achieve a higher accuracy than LIBSVM, even if only one worker is employed. This indicates that humans are much better at natural language understanding than machines. For such tasks, if high accurate results are required, crowdsourcing is a promising approach.

[2]http://www.csie.ntu.edu.tw/~cjlin/libsvm/

### 5.1.2 Accuracy Analysis

In TSA, we first apply Theorem 1 to estimate the number of workers required. This is a conservative estimation. To reduce cost, binary search is used to refine the estimation. Figure 6 compares the conservative estimation with the refined estimation generated by the binary search. We change the user required accuracy from 0.65 to 0.99 and find that the refined estimation is less than half of the conservative estimation. In the remaining experiments, we use the refined estimation to determine the number of workers required for each HIT.

We next present the accuracy for the three verification models, namely *Half-Voting*, *Majority-Voting* and our proposed *Probability-based Verification* model. Figure 7 shows that when the number of workers increases, we can get a higher accuracy. Among the three verification models, our probability-based approach achieves a much higher accuracy than the other two. When 29 workers are employed, the probability-based model improves the accuracy to 0.99. This verifies the benefit of considering workers' historical performance.

We proceed to investigate the effectiveness of the three verification models with respect to a user required accuracy. Figure 8 shows the result. When the requester specifies a required accuracy, TSA estimates the number of workers needed to achieve that accuracy. The real accuracy is computed by comparing the workers' answers with the ground truth. The red line in the figure denotes the user required accuracy. We observe that the probability-based verification model always provides a satisfactory result while the results of the other two models are below the required accuracy in most cases.

We can observe that the accuracy of the *Half-Voting* model is worse than our estimation. The reason is as follows. First, the estimated number of workers ties to users' mean accuracy. The mean accuracy used in the prediction model is an overall accuracy, which is collected across various questions. However, for some difficult questions, workers' accuracies could be much lower. As a result, the number of workers needed in voting models is more than the estimated number. For example, the following tweet about movie *The Last Airbender* expresses a positive opinion whereas most workers classify it into the negative category because of the word "sucks".

> My nephew just said that Avatar: The Last Airbender sucks... I'm disowning him.

The second reason can be explained based on the results of Figure 9 and Figure 10. Figure 9 shows the percentage of tweets with no answers in the two voting-based models. In some cases, the *Half-Voting* and *Majority-Voting* models fail to provide a result as none of the answers is discriminative (All answers get no more than half votes or more than one answers get the same number of votes). When the number of workers increases, *Majority-Voting* can solve the tie more easily. However, for the *Half-Voting* strategy, there are still about 15% of the tweets that cannot obtain answers with more than half the amount of votes. In Figure 10, when we vary the number of tweet reviews, we observe that the percentage of no-answer reviews is fairly stable. This phenomenon indicates that the reviews with non-discriminative answers are almost uniformly distributed among all reviews.

### 5.1.3 Online Processing

One advantage of our crowdsourcing engine is its ability in supporting online processing. It can provide an approximate result without waiting for all the workers to finish their jobs. Specifically, TSA will generate an initial result as soon as the first answer is returned. Then it will gradually refine the results as more answers arrive until the termination condition is satisfied. This allows us to terminate a HIT and cap the processing cost[3].

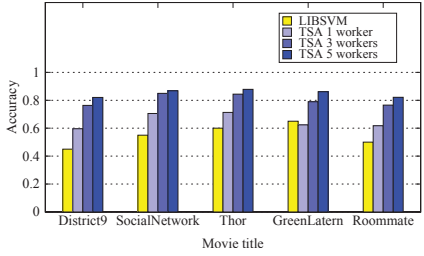

**Figure 5:** Crowdsourcing vs. SVM Algorithm

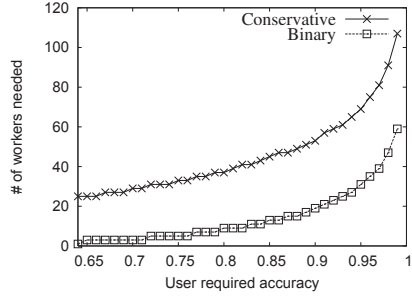

**Figure 6:** Number of Workers Required

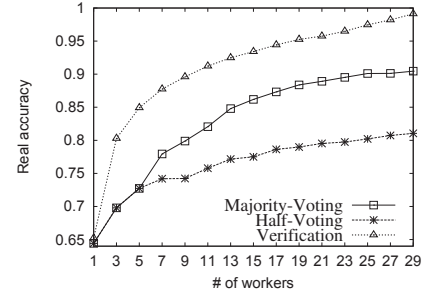

**Figure 7:** Accuracy Comparison wrt. Number of Workers

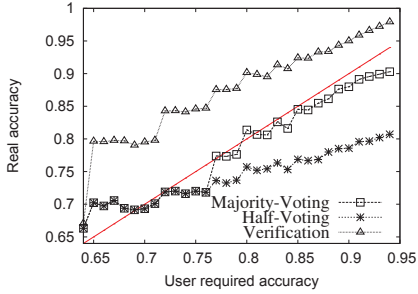

**Figure 8:** Accuracy Comparison wrt. User Required Accuracy

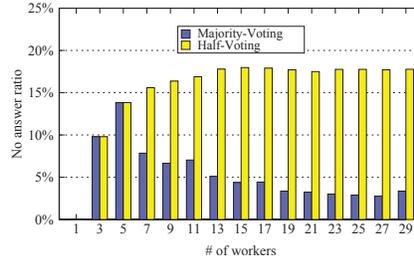

**Figure 9:** Percentage of No-Answer Reviews wrt. Number of Workers

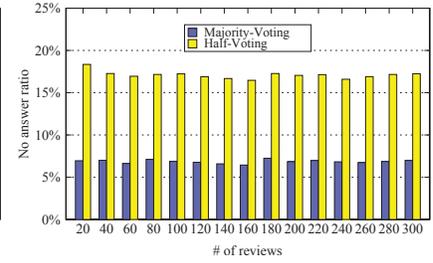

**Figure 10:** Percentage of No-Answer Reviews wrt. Number of Reviews

One interesting observation in our experiments is that the accuracy of the approximate result varies significantly for different answer arriving sequences. Figure 11 shows the accuracy of the same HIT under four different answer sequences. The red line is the user-required accuracy 0.94. Sequence 4 results in a low starting accuracy because the first two workers of sequence 4 provide incorrect answers. Therefore, in online processing, we must update the confidence of the current result dynamically based on the answers received as early termination may potentially degrade the accuracy.

We evaluate the three termination strategies as discussed in Section 4.2.2. Figure 12 shows the effect of early termination on the number of workers. The red line denotes the estimated number of workers via our refined prediction model. The *MinMax* strategy generates the most conservative estimation, but it still reduces the number of workers by 20%. The *ExpMax* strategy is the most aggressive one, which can save more than 50% of workers. In Figure 13, we show the accuracies of the different termination strategies. The x-axis is the accuracy requirement specified by the user and the y-axis is the real accuracy measured against the ground truth. We can see that the *MinMax* and *ExpMax* strategies satisfy the user required accuracy (denoted as red line) in all cases while *MinExp* fails to meet the requirement at a few points. In view of the need for reducing the number of workers while maintaining good accuracy, we propose to adopt the *ExpMax* termination strategy.

---

[3]In AMT, we can cancel a HIT when we detect that the answers are good enough. By doing so, we do not need to pay workers who have yet submitted their answers.

### 5.1.4 Effect of Sampling

TSA verifies the answers using the probability-based verification model, which relies on workers' historical performance. The AMT system records an approval rate for each worker, which implies his accuracy in general. However, the workers' approval rates are not public due to privacy concerns. To collect the statistics, we publish 500 HITs requiring workers to fill in their approval rate. We also compute the workers' accuracies of answering TSA queries. We observe the distribution of their approval rate in AMT is very different from that of real accuracy in TSA, as shown in Figure 14. The reasons are two-fold. On one hand, there are various types of tasks in AMT and it is natural that people cannot be experts in all domains. On the other hand, some requesters set automatic approval for all workers without checking the answers. This results in a high average approval rate in AMT. Therefore, we adopt a sampling approach to estimate workers' accuracy.

Given $n$ works, we compute their accuracies $A^j = \{a_1^j, a_2^j, ..., a_n^j\}$ under a sampling rate $j\%$. We vary the sampling rate and plot the mean accuracy $\mu^j$ and average absolute error $err^j$ in Figure 15, where $\mu^j$ and $err^j$ are defined as follows:

$$\mu^j = \frac{1}{n}\sum_{i=1}^{n} a_i^j, \ \ err^j = \frac{1}{n}\sum_{i=1}^{n} |a_i^j - a_i^{100}|$$

As shown, both mean accuracy and average error are stable when the sampling rate is higher than 10%. More precisely, mean accuracy remains nearly constant and average error approaches 0.

We also study the effect of sampling rate on accuracy in our verification model. Figure 16 plots the result. We vary the sampling rate from 5%, 10% to 20% and compare the result to 100%-sampling accuracy. The red line represents the user required accuracy. We can see that the verification has a better accuracy with a higher sampling rate. When the user required accuracy is lower than 0.75, all sampling rates are satisfactory. The result meets all of the user required accuracy only with a sampling rate no less than 20%. Moreover, the accuracy under 20% sampling rate has only a small gap compared to that under 100% sampling. We use 20% sampling rate in all of our verification experiments.

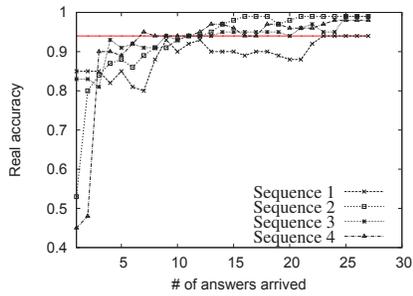

**Figure 11:** Effect of Answer Arriving Sequence

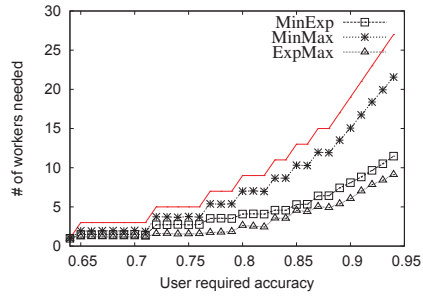

**Figure 12:** Effect of Early Termination on Worker Number

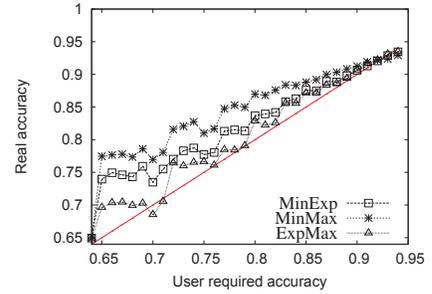

**Figure 13:** Effect of Early Termination on Accuracy

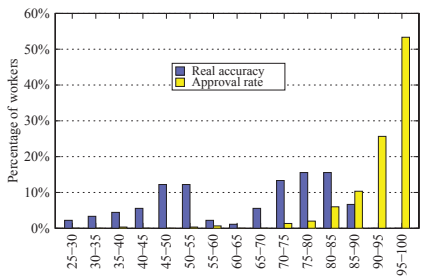

**Figure 14:** Worker Accuracy vs. Approval Rate

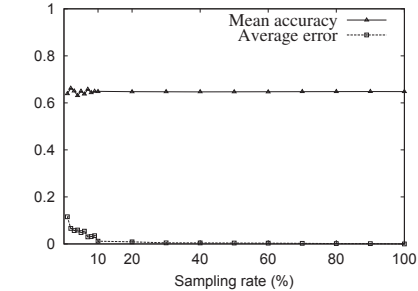

**Figure 15:** Effect of Sampling Rate on Worker Accuracy

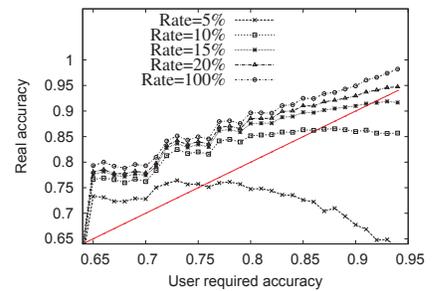

**Figure 16:** Effect of Sampling Rate on Verification Accuracy

## 5.2 Application 2: IT

In this experiment, we evaluate our model in the context of image tagging application. We use 100 Flicker images as our queries. For each image, we give a set of candidate tags and let 30 workers to choose the related ones. The candidate tags include Flicker tags and some embedded noise tags.

Again, we first show the advantages of crowdsourcing over the applications on dealing with image tagging task. We compare our result with ALIPR[4]. ALIPR[13] is an automatic image annotation system which applies 2-D Hidden Markov model and clustering techniques. The accuracy comparison result is shown in Figure 17. We use 5 groups of images. Each group contains top 20 Flicker images returned by a tag. The figure clearly shows the accuracy gap between ALIPR and crowdsourcing approach. ALIPR achieves its best accuracy 30% on tag `sun` and has only 12.6% accuracy on tag `apple`, whereas in our crowdsourcing system, we can reach more than 80% even with only one worker employed.

We next study the effectiveness of our model. Recall that our model first estimates the number of workers for a specified accuracy requirement and then applies a probability-based model to verify the result. Figure 18 shows the accuracy achieved with respect to the user required accuracy. As before, the red line denotes the user required accuracy. It can be seen from the figure that our model can always satisfy user's requirement.

---

[4]http://alipr.com/

## 6. RELATED WORK

The emergence of Web 2.0 systems has significantly increased the applicability and usefulness of crowdsourcing techniques. A complex job can be split into many small tasks and assigned to different online workers. Amazon's AMT and CrowdFlower[5] are popular crowdsourcing platforms. Studies show that users exhibit different behaviors in such micro-task markets [11]. A good incentive model is required in task design [10].

Recently, crowdsourcing has been adopted in software development. Instead of answering all requests with computer algorithms, some human-expert tasks are published on crowdsourcing platforms for human workers to process. Typical tasks include image annotation [21][18], information retrieval [1][8] and natural language processing [3][12][17]. These are tasks that even state-of-the-art technologies cannot accomplish with satisfactory accuracy, but could be easily and correctly done by humans.

Crowdsourcing techniques have also been introduced into the database design. Qurk [16][15] and CrowdDB [6] are two examples of databases with crowdsourcing support. In these database systems, queries are partially answered by AMT platform. Our system, CDAS, adopts a similar design. On top of the crowdsourcing database, new query languages, such as hQuery [20], have been proposed, which allows users to exploit the power of crowdsourcing. Other database applications, such as graph search [19], can be enhanced with crowdsourcing techniques as well.

One main obstacle that prevents enterprise-wide deployment of crowdsourcing-based applications is quality control. Human workers' behaviors are unpredictable, and hence, their answers may be arbitrarily bad. To encourage them to provide high-quality answers, monetary rewards are required. Munro et al. [17] showed how to design a good incentive model to optimize workers' participation and contributions. Ipeirotis et al. [9] presented a scheme to rank the qualities of workers while Ghosh et al. [7] tried to accurately identify abusive content. Unlike previous efforts, in this paper, we have designed a feasible model that balances monetary cost and accuracy, and proposed a crowdsourcing query engine with quality control. One of the main challenges of our query engine is how to integrate the conflicting results of human workers. The similar problem has been well studied in the data fusion systems, for examples [4][14]. We extended the models proposed in [4][14] to select and verify the crowdsourcing results in our CDAS.

---

[5]http://crowdflower.com/

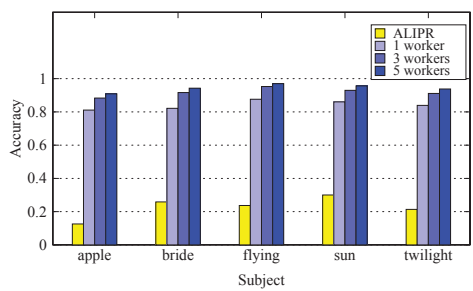

**Figure 17:** Crowdsourcing vs. ALIPR

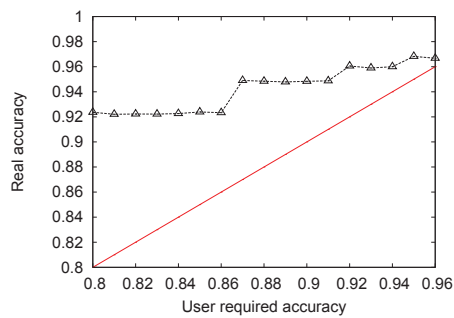

**Figure 18:** Accuracy Obtained wrt. User Required Accuracy

## 7. CONCLUSION

Crowdsourcing techniques allow application developers to harness the natural expertise of human workers to perform complex tasks that are very challenging for computers. However, as humans are prone to errors, there is no guarantee for the results of crowdsourcing. In this paper, we introduced the quality-sensitive answering model in our Crowdsourcing Data Analytics System, CDAS. The model guides the query engine to generate proper query plans based on the accuracy requirement. It consists of two sub-models, the prediction model and the verification model. The prediction model estimates the number of workers required for a specific task while the verification model selects the best answer from all returned ones. To improve users' experience, when verifying the results, our model embraces online processing techniques to update answers gradually. By adopting the models, CDAS can provide high-quality results for different crowdsourcing jobs. In this paper, we have implemented a twitter sentiment analytics job and an image tagging job on CDAS. We used real Twitter data and Flickr data as our queries. Amazon Mechanical Turk was employed as our crowdsourcing platform. The results show that our proposed model can provide high-quality answers while keeping the total cost low.

## 8. ACKNOWLEDGEMENT

The work of this paper was in part supported by Singapore MDA grant R-252-000-376-279.